\documentclass[aps,prl,twocolumn,showpacs,floatfix,superscriptaddress]{revtex4}
\usepackage{color,graphicx}
\usepackage{hyperref}
\begin{document}
\flushbottom

\title{Magnetic field dependence of the density of states in the multiband superconductor $\beta$-Bi$_2$Pd}

\author{E. Herrera}
\affiliation{Laboratorio de Bajas Temperaturas, Departamento de F\'isica de la Materia Condensada, Instituto Nicol\'as Cabrera and Condensed Matter Physics Center (IFIMAC), Universidad Aut\'onoma de Madrid, E-28049 Madrid, Spain}

\author{I. Guillam\'on}
\affiliation{Laboratorio de Bajas Temperaturas, Departamento de F\'isica de la Materia Condensada, Instituto Nicol\'as Cabrera and Condensed Matter Physics Center (IFIMAC), Universidad Aut\'onoma de Madrid, E-28049 Madrid, Spain}
\affiliation{Unidad Asociada de Bajas Temperaturas y Altos Campos Magn\'eticos, UAM, CSIC, E-28049 Madrid, Spain}

\author{J.A. Galvis}
\affiliation{Laboratorio de Bajas Temperaturas, Departamento de F\'isica de la Materia Condensada, Instituto Nicol\'as Cabrera and Condensed Matter Physics Center (IFIMAC), Universidad Aut\'onoma de Madrid, E-28049 Madrid, Spain}
\affiliation{Departamento de ciencias naturales, Facultad de ingenieria, Universidad Central, Bogot\'a, Colombia.}

\author{A. Correa}
\affiliation{Instituto de Ciencia de Materiales de Madrid, Consejo Superior de Investigaciones Cient\'{\i}ficas (ICMM-CSIC), Sor Juana In\'es de la Cruz 3, 28049 Madrid, Spain.}

\author{A. Fente}
\affiliation{Laboratorio de Bajas Temperaturas, Departamento de F\'isica de la Materia Condensada, Instituto Nicol\'as Cabrera and Condensed Matter Physics Center (IFIMAC), Universidad Aut\'onoma de Madrid, E-28049 Madrid, Spain}

\author{R.F. Luccas}
\affiliation{Instituto de Ciencia de Materiales de Madrid, Consejo Superior de Investigaciones Cient\'{\i}ficas (ICMM-CSIC), Sor Juana In\'es de la Cruz 3, 28049 Madrid, Spain.}

\author{F. J. Mompean}
\affiliation{Instituto de Ciencia de Materiales de Madrid, Consejo Superior de Investigaciones Cient\'{\i}ficas (ICMM-CSIC), Sor Juana In\'es de la Cruz 3, 28049 Madrid, Spain.}
\affiliation{Unidad Asociada de Bajas Temperaturas y Altos Campos Magn\'eticos, UAM, CSIC, E-28049 Madrid, Spain}

\author{M. Garc{\'i}a-Hern{\'a}ndez}
\affiliation{Instituto de Ciencia de Materiales de Madrid, Consejo Superior de Investigaciones Cient\'{\i}ficas (ICMM-CSIC), Sor Juana In\'es de la Cruz 3, 28049 Madrid, Spain.}
\affiliation{Unidad Asociada de Bajas Temperaturas y Altos Campos Magn\'eticos, UAM, CSIC, E-28049 Madrid, Spain}

\author{S. Vieira}
\affiliation{Laboratorio de Bajas Temperaturas, Departamento de F\'isica de la Materia Condensada, Instituto Nicol\'as Cabrera and Condensed Matter Physics Center (IFIMAC), Universidad Aut\'onoma de Madrid, E-28049 Madrid, Spain}
\affiliation{Unidad Asociada de Bajas Temperaturas y Altos Campos Magn\'eticos, UAM, CSIC, E-28049 Madrid, Spain}

\author{J.P. Brison}
\affiliation{Univ. Grenoble Alpes, INAC-SPSMS, F-38000 Grenoble, France}
\affiliation{CEA, INAC-SPSMS D-38000 Grenoble, France}

\author{H. Suderow}
\email[Corresponding author: ]{hermann.suderow@uam.es}
\affiliation{Laboratorio de Bajas Temperaturas, Departamento de F\'isica de la Materia Condensada, Instituto Nicol\'as Cabrera and Condensed Matter Physics Center (IFIMAC), Universidad Aut\'onoma de Madrid, E-28049 Madrid, Spain}
\affiliation{Unidad Asociada de Bajas Temperaturas y Altos Campos Magn\'eticos, UAM, CSIC, E-28049 Madrid, Spain}

\date{\today}

\begin{abstract}
We present very low temperature scanning tunneling microscopy (STM) experiments on single crystalline samples of the superconductor $\beta$-Bi$_2$Pd. We find a single fully isotropic superconducting gap. However, the magnetic field dependence of the intervortex density of states is higher than the one expected in a single gap superconductor, and the hexagonal vortex lattice is locked to the square atomic lattice. Such increase in the intervortex density of states and vortex lattice locking have been found in superconductors with multiple superconducting gaps and anisotropic Fermi surfaces. We compare the upper critical field $H_{c2}(T)$ obtained in our sample with previous measurements and explain available data within multiband supercondutivity. We propose that $\beta$-Bi$_2$Pd is a single gap multiband superconductor. We anticipate that single gap multiband superconductivity can occur in other compounds with complex Fermi surfaces.\end{abstract}


\maketitle

\section{Introduction}

Superconductivity is often found in binary metallic compounds with critical temperatures of the order of liquid helium temperature. Some are reviewed in Ref.\cite{Matthias63} and give type II superconductors. Among them, MgB$_2$ is peculiar, with a critical temperature $T_c$ = 40K unsurpassed by related binary compounds. Such a high T$_c$ results from the combination of the strong electron-phonon coupling of the two-dimensional $\sigma$ bands and weak interband mixing with the three dimensional $\pi$ bands\cite{Liu01,Rubio01,Bascones01,Giubileo01,Szabo01,Iavarone02,Zehetmayer13}. Each set of bands is derived from orthogonal orbital wavefunctions, leading to two well defined superconducting gap features. Multigap and multiband superconductivity have been conceptually linked together since the discovery of MgB$_2$, suggesting that materials showing different Fermi surface sheets also have different superconducting gaps in each sheet.

Under magnetic fields, vortex core overlap is governed by the Fermi surface velocity (see Refs.\cite{Hohenberg67,Gurevich07,Hirschfeld11}). In MgB$_2$, enhanced vortex core overlap has been observed and related to the Fermi surface properties of the sheet having a smaller sized superconducting gap\cite{Eskildsen02,Eskildsen03,Kohen05,Zehetmayer13}. This leads to a strong increase of the density of intervortex quasiparticle excitations measured by Scanning Tunneling Microscopy and Spectroscopy (STM)\cite{Eskildsen02,Eskildsen03}, and of the overall density of states measured by specific heat or thermal conductivity, when applying a magnetic field. The upper critical field shows a positive curvature, instead of the negative curvature expected within single gap s-wave BCS superconductivity\cite{Werthamer66,Shulga98,Imai12,Suderow04b,Suderow05d,Tissen13,Larbalestier01,Sologubenko02,Gurevich07}. The same occurs in many different compounds, including heavy fermions, borocarbides, and Fe based superconductors\cite{Bauer01,Schmiedeshoff01,Lipp02,Bouquet02,Kacmarcik10,Seyfarth05,Seyfarth06,Hirschfeld11}.  
So far, the increased density of states and positive curvature of $H_{c2}$ close to $T_c$ have been explained through multigap superconductivity. However, the detailed mixed phase properties of a single gap s-wave superconductor and the possible influence of multiband and of anisotropic Fermi surfaces in the mixed phase remain unclear.

\begin{figure}[ht]
\includegraphics[width=0.45\textwidth]{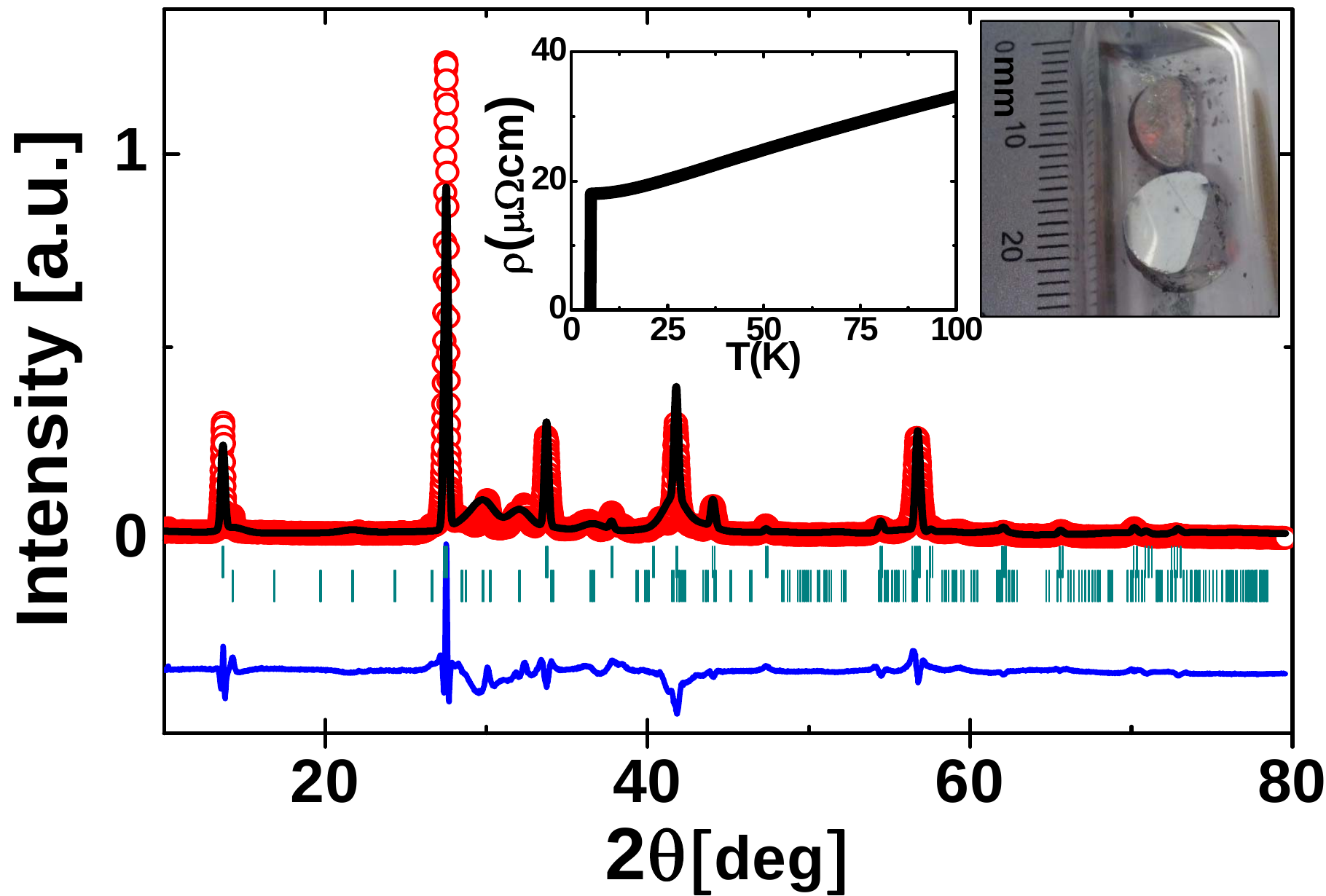}
\caption{Powder diffraction pattern of $\beta$-Bi$_2$Pd. Red symbols are the experimental points. The black line is the best fit to $\beta$-Bi$_2$Pd diffraction pattern\protect\cite{FullProf}. Residuals are given by the blue line. The two series (upper and lower) of vertical green strikes represent, respectively, the position in 2$\theta$ scale of the reflections from the $\beta$-Bi$_2$Pd (I4/mmm) and $\alpha$-Bi$_2$Pd ($C 12/m1$) phases. Insets show a photograph of one $\beta$-Bi$_2$Pd crystal, and the temperature dependence of the resistivity.}
\end{figure}

Here we show how superconducting features are modified by a multiband Fermi surface. We measure $\beta-$Bi$_2$Pd (T$_c$ = 5 K \cite{Alekseevski54,Imai12}) with a very low temperature STM. Our sample is moderately in the dirty limit (mean free path $\ell$ smaller than superconducting coherence length $\xi$, $\ell<\xi$) and the Fermi surface shows multiple sheets of mixed orbital character\cite{Shein12,Imai12,Sakano15}. The situation is opposed to MgB$_2$, with $\sigma$ and $\pi$ sheets that remain well separated even in presence of defects\cite{Gurevich07}. At 150 mK, we obtain atomic scale imaging and the hexagonal vortex lattice. At zero field we find single gap behavior following s-wave BCS theory but observe multiband properties in the mixed phase. The density of states in-between vortices increases more than expected for an isotropic single gap superconductor and the hexagonal vortex lattice locks to the square crystal lattice.

\section{Crystal growth and experimental methods}

Single crystals of $\beta$-Bi$_2$Pd were grown using slight excess of Bi \cite{Canfield1992,CanfieldBook}. We grew our samples from high purity Bi (Alfa Aesar 99.99 $\%$) and Pd (Alfa Aesar 99.95 $\%$). Bi and Pd were introduced in quartz ampoules and sealed at 140 mbar of He gas. Then, ampoules were heated from room temperature to 900 $^\circ$C in 3 h, maintained 24 h at this temperature, slowly cooled down to 490 $^\circ$C in 96 h and finally cooled down to 395 $^\circ$C in 200 h. This temperature is about 15 $^\circ$C above the temperature for the formation of the $\alpha$-Bi$_2$Pd phase\cite{Okamoto94}. To avoid formation of the $\alpha$ phase, we quenched the crystals down to ambient temperature by immersion in cold water. We obtained large crystals of 5 mm $\times$ 5mm $\times$ 3mm. To characterize them, we made x-ray diffraction on crystals milled down to powder (Fig.\ 1, using x rays with wavelength $1.54$ $\AA$). We find $\beta$-Bi$_2$Pd (I4/mmm, see Ref.\cite{Zhuravlev57}) with refined lattice parameters $a = b = 3.36(8)$ \AA\ and $ c=12.97(2)$ \AA\  and no trace of $\alpha$-Bi$_2$Pd. We made in total twelve growths, varying slightly the conditions for the quench, growth temperature and initial composition, and obtained always crystals with a resistivity vs temperature very similar to the one shown in Fig.\ 1. The temperature dependence of the resistivity is shown in the inset of Fig.\ 1. The superconducting transition in our sample of $\beta$-Bi$_2$Pd crystals occurs at 5 K. Previous resistivity measurements in this material reported a slightly higher value of T$_c$ (around 0.3 K larger) and a residual resistivity three time smaller than the one found here\cite{Imai12}. The specific heat transition of our samples is sharp, of about 30 mK width\cite{SamuelyToBePublished}, contrasting the transition width of about 300 mK reported in Ref.\cite{Imai12}. We measured also the upper critical field $H_{c2}(T)$ using resistivity and susceptibility as a function of temperature or magnetic field for the field applied parallel to the c-axis. The results coincide with the positions where we also observed vanishing superconducting features in STM tunneling conductance.

To make the STM measurements, we use a home built set-up installed in a dilution refrigerator with an energy resolution in the tunneling spectroscopy of 0.15 K. Construction is similar to Ref.\cite{Suderow11}. We also took a few tunneling data at fields parallel to the surface using a three axis coil system described in Ref.\cite{Galvis15}. We use an Au tip cleaned by repeated indentation on an Au sample as described in Ref.\cite{Rodrigo04b}. We make the STM measurements in a sample roughly one mm thick, which was cleaved using a scotch tape at ambient conditions after glueing it using silver epoxy to the sample holder. Usually, bias voltage is kept at 10 mV or below, and the tunneling conductance is of a few tenths of $\mu$S. Topography and vortex lattice images are independent of tunneling parameters. To obtain vortex lattice images we cut the feedback loop at each point and make full I-V curves, as in previous work\cite{Guillamon08c}. No filtering or image treatment is applied to the topography and conductance maps shown here.

\begin{figure}
\includegraphics[width=0.45\textwidth]{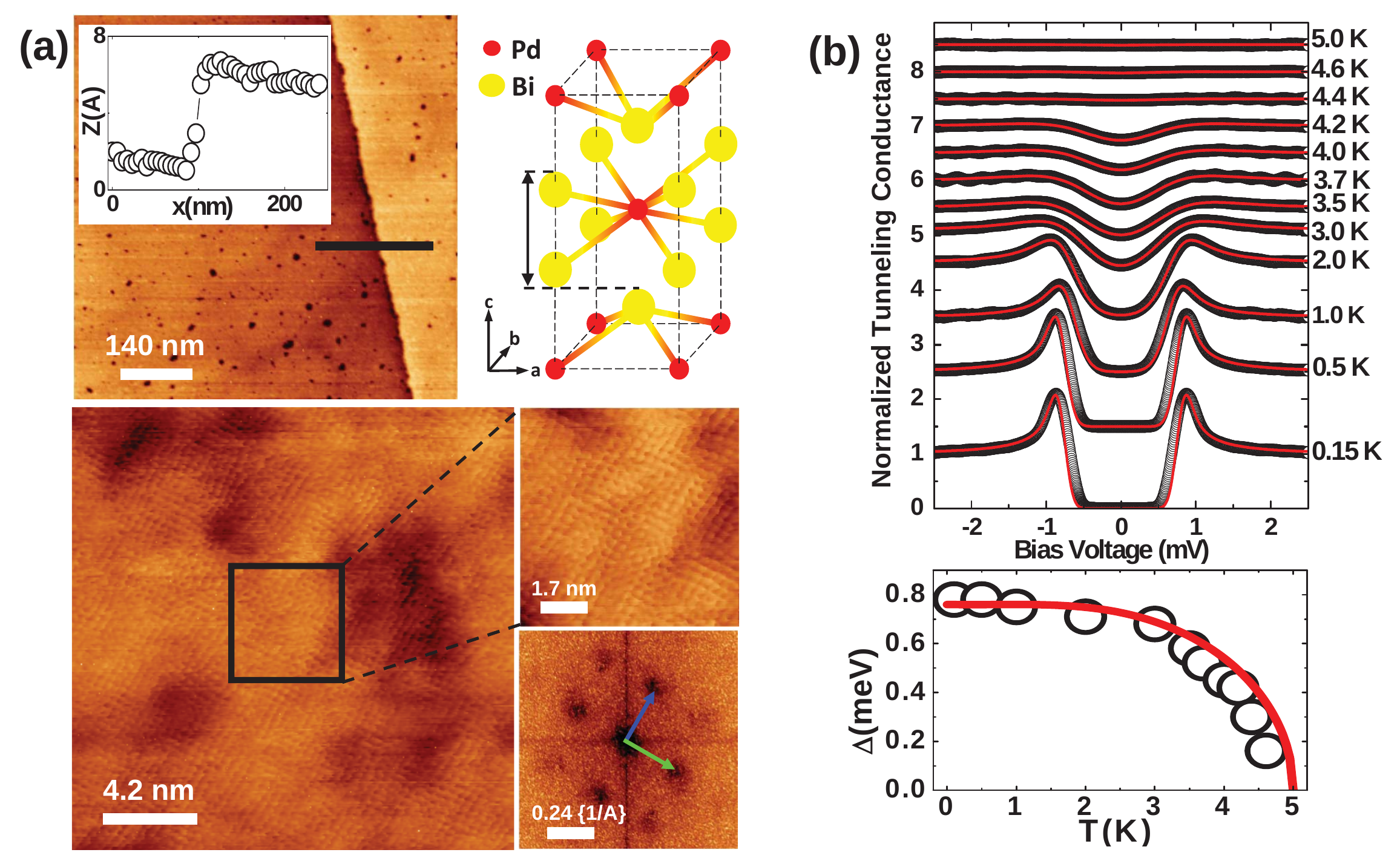}
\caption{a) Atomic scale imaging of the surface of $\beta$-Bi$_2$Pd at different length scales. Images are taken at a bias voltage of 10 mV and conductance 1$\mu$S at 0.15 K. Lattice structure, highlighting the cleaving plane with distance between two successive planes $d=6.6$\AA\ is also shown. The inset in the top left image shows a cut through a line showing a jump whose size corresponds to the distance between cleaving planes. Bottom right panel is a Fourier transform of the atomic size images. Arrows give crystalline axis. b) Temperature dependence of the experimental tunneling conductance (black dots). Red lines are fits to the s-wave BCS expression at each temperature, leaving $\Delta$ as the only free parameter. The values of $\Delta$ obtained are plotted in the bottom panel, together with the temperature dependence obtained from BCS theory (red line).}
\end{figure}

\section{Results}

\begin{figure*}
\includegraphics[width=0.9\textwidth]{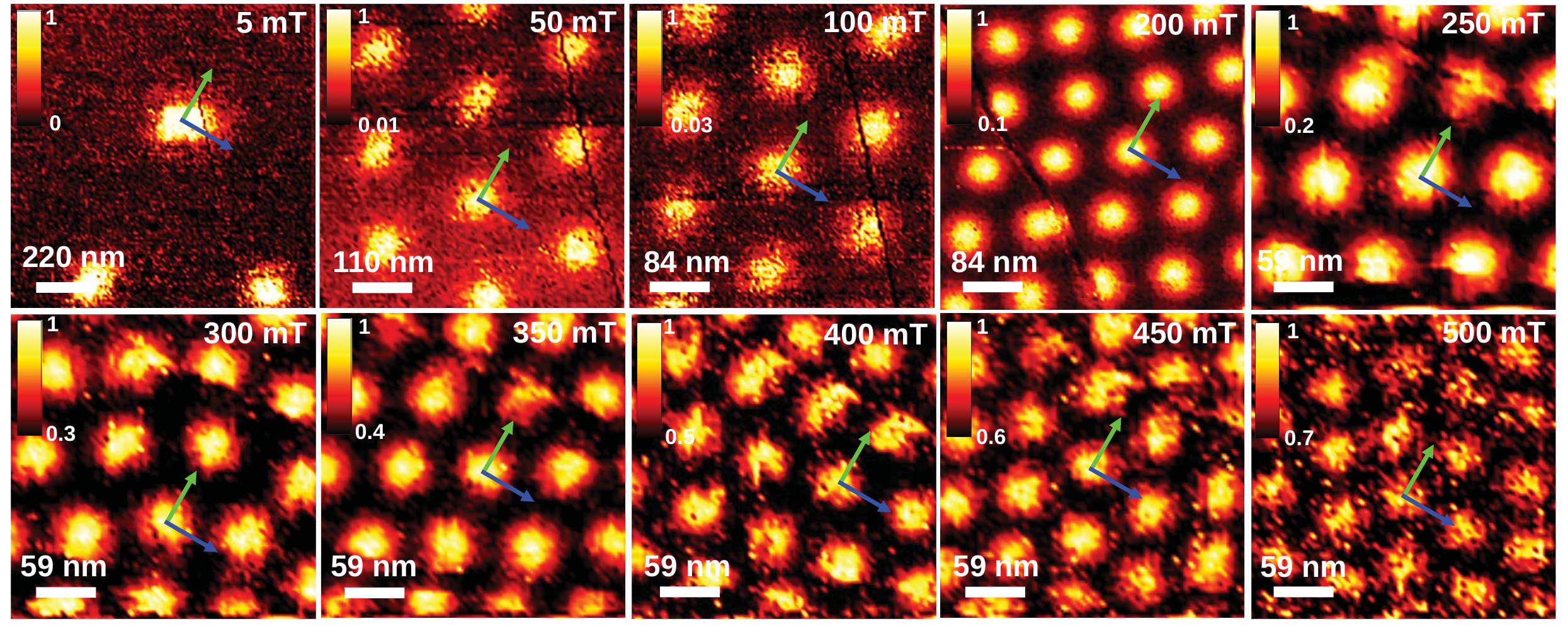}
%\vskip -5cm
\caption{The vortex lattice as a function of the magnetic field applied parallel to the c axis direction of $\beta$-Bi$_2$Pd. The vortex lattice has a hexagonal symmetry for all values of the magnetic field. The images were taken at 150 mK and the magnetic field was increased up to H$_{c2}=$0.6 T. Constrast in the zero bias conductance is shown with color scales. The orientation of the crystalline axis is shown as green and blue arrows in each panel.}
\end{figure*}

The tetragonal structure with Bi-Pd blocks (Fig.\ 2) suggests that it is easy to obtain clean and atomically flat surfaces by cleavage or exfoliation. In Fig.\ 2a we show atomic resolution topography images taken on the surface of the sample at 0.15 K. We find indeed a square atomic lattice. The Fourier transform of the topography images gives a lattice parameter of $a=b=3.3$\ \AA, coinciding with crystal structure values. Surfaces are atomically flat over hundreds of nm. Small steps are sometimes viewed in the images. The step in Fig.\ 2a (top left panel and inset) is of 6.5\ \AA\ height, which corresponds to the distance between adjacent Bi-Pd groups (black arrow in top right panel of Fig.\ 2a). Bi-Pd groups are more strongly coupled than the Bi-Bi sheets, because the directional Bi-Bi bonds are weaker than the Bi-Pd bonds\cite{Shein12}. Thus, we conclude that the surfaces in Fig.\ 2a are made out of the square Bi lattice.

Fig.\ 2b shows the tunneling conductance vs bias voltage as a function of temperature. At 0.15 K, we find clear superconducting quasiparticles peaks and no conductance at zero and low bias. We can fit our data using single gap BCS theory and $\Delta=0.76$ meV. Temperature dependence of the superconducting features shows that superconductivity disappears at about 5 K.

When we apply a magnetic field parallel to the $c$ axis, we observe an ordered hexagonal vortex lattice on large atomically flat regions. In Fig.\ 3 we show vortex lattice images from 5 mT to 500 mT. A hexagonal Abrikosov lattice is observed in all images. The intervortex distance is modified as expected for the hexagonal vortex lattice $d_{\Delta}=1.075 \sqrt{\frac{\phi_0}{B}}$ (with $\phi_0$ being the flux quantum).

We find that the orientation of the vortex lattice is determined by the underlying crystalline lattice (arrows in Fig.\ 3). One of the three main vortex lattice directions is always parallel to one of the two crystalline axes. This gives two equivalent orientations for the hexagonal vortex lattice at any magnetic field.

\begin{figure}[ht]
\includegraphics[width=0.45\textwidth]{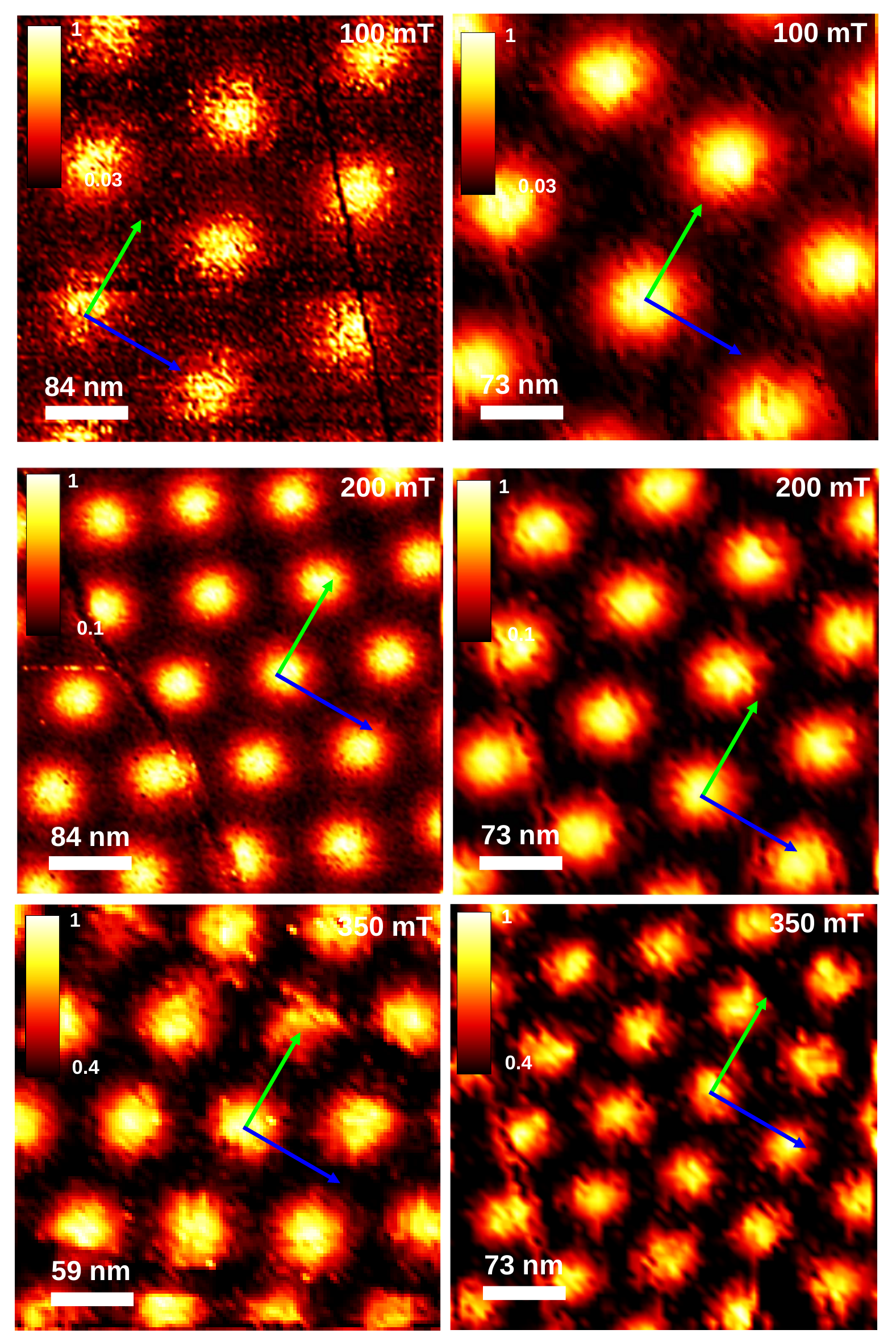}
\vskip -0.2cm
\caption{Vortex lattice at three different magnetic fields. In each magnetic field we show two vortex images obtained at different scanning windows. Note the change in the orientation with respect to the atomic lattice, as marked by green and blue arrows. Contrast (zero bias conductance) is shown with the color scales.}
\end{figure}

At a fixed magnetic field, we find the two different vortex lattice orientations when changing the scanning windows (Fig.\ 4), obtained by moving in-situ the sample holder using the method described in Ref.\cite{Suderow11}. Thus, the orientation of the hexagonal vortex lattice forms domains oriented along one crystal axis. The size of these domains is considerably larger than the scanning window (2 $\mu$m $\times$ 2 $\mu$m).

\begin{figure*}
\includegraphics[width=0.9\textwidth]{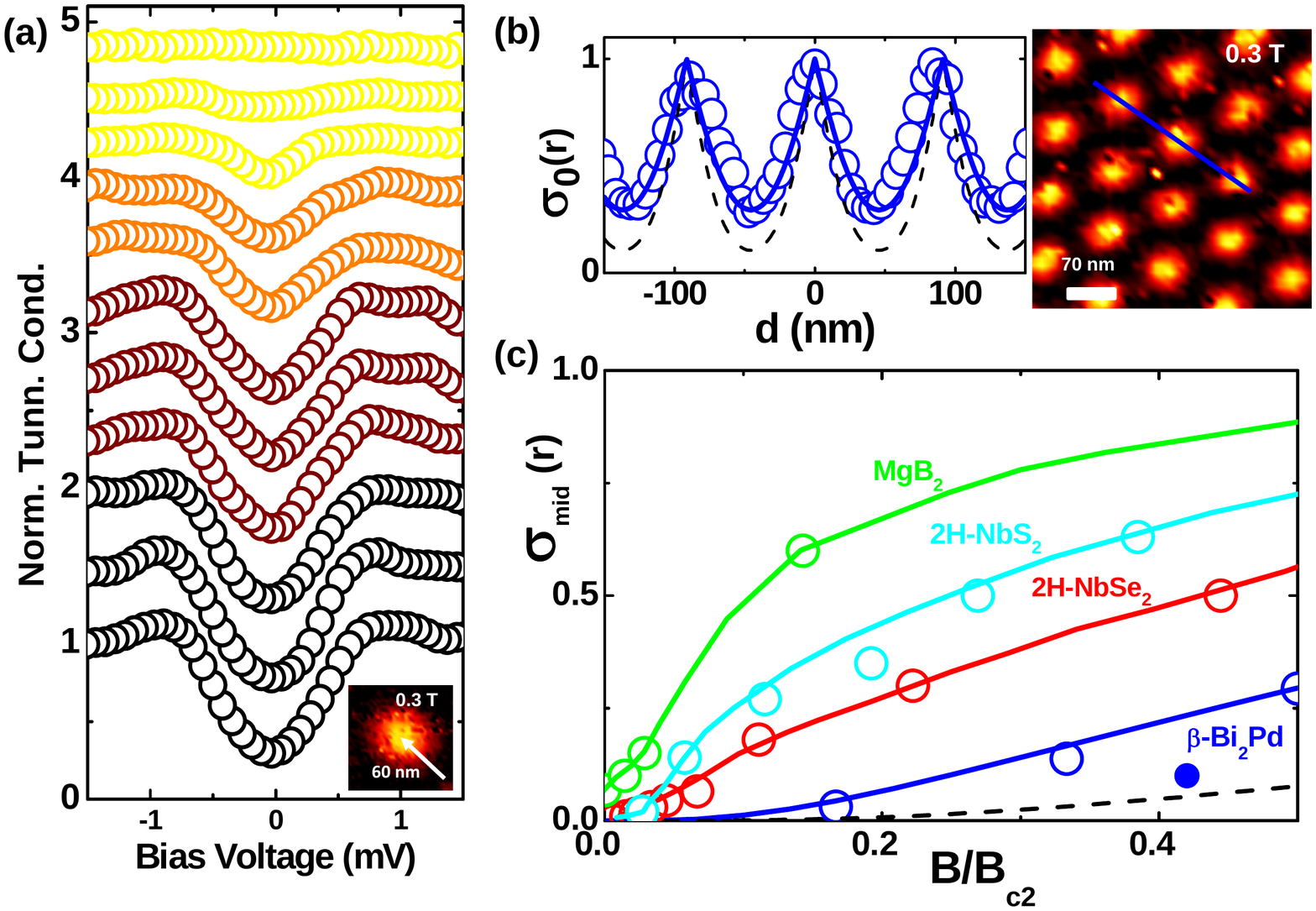}
\caption{a) The tunneling conductance vs bias voltage when entering a vortex core at 0.3 T along the path shown by the white arrow in the inset (one curve each 6 nm approximately). In b) we show the normalized zero bias conductance $\sigma_0$ as a function of the position at 150 mK and at $H/H_{c2}=0.5$ (left panel, data shown as points, continuous blue line is a guide to the eye and black dashed line is calculated as discussed in the text). c) we show the magnetic field dependence of the zero bias tunneling conductance $\sigma_{mid}$ exactly in-between vortices (blue points, continuous blue line is a guide to the eye) when the field is applied along the c-axis (open blue points) and when the field is applied along the plane (filled blue point). Data from 2H-NbSe$_2$, 2H-NbS$_2$ and MgB$_2$ (points, lines are guides to the eye) are Refs.\protect\cite{Eskildsen02,Eskildsen03,Guillamon08PRB,Guillamon08c}.}
\end{figure*}

The spatial dependence of the superconducting density of states in and around vortices shows that the gap fully closes inside the vortex core(Fig.\ 5a). We do not observe signatures of Caroli-de Gennes-Matricon Andreev core states \cite{Caroli64,Guillamon08PRB,Hess90}. From the residual resistivity of our samples ($\rho=18\  \mu \Omega cm$ just above T$_c$) we estimate the mean free path using Drude formula and find $\ell = 15.3$ nm. On the other hand, the in-plane coherence length from the upper critical field (discussed below, H$_{c2}(T = 0 K)=\frac{\Phi_0}{2\pi\xi^2}$) yields $\xi=$ 23 nm. Thus, $\ell<\xi$ and Caroli-de Gennes-Matricon states are smeared by defect scattering\cite{Renner91}.

\begin{figure}[ht]
\includegraphics[width=0.45\textwidth]{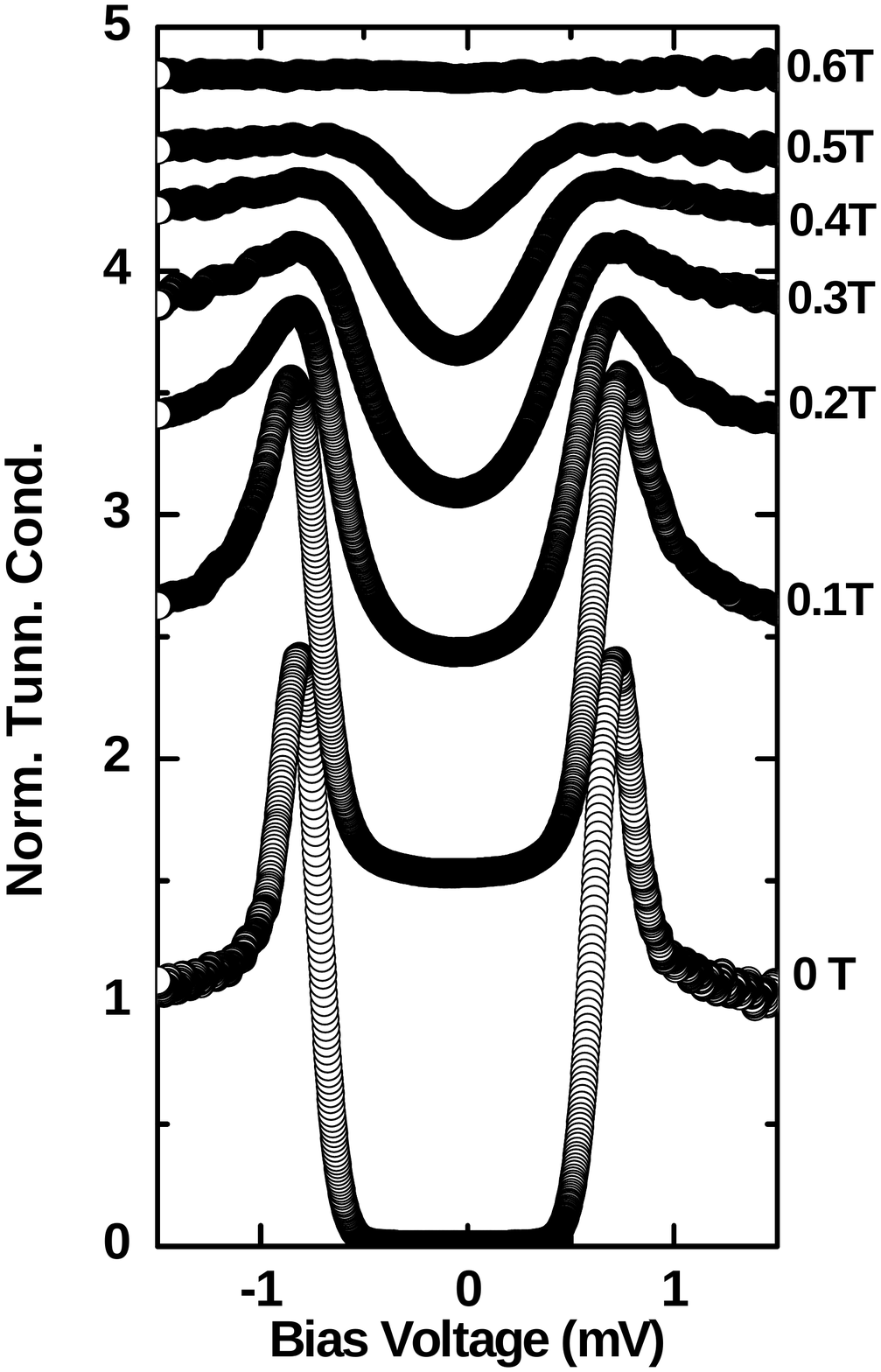}
\vskip -0.5cm
\caption{Magnetic field dependence of the full normalized tunneling conductance $\sigma_{mid}(V)$ obtained exactly at the center between vortex cores. Magnetic field is applied along the c-axis. Curves are shifted vertically for clarity.}
\end{figure}

In Fig.\ 5b we show the spatial dependence of the zero bias conductance $\sigma_0(r)$ along a path crossing several vortex cores at 0.3 T in $\beta-$Bi$_2$Pd (blue points), and the expectation assuming vortex core overlap far from H$_{c2}$. The latter is calculated by summing over relevant neighbors ($r_i$) using  $\sigma_0(r)=\sum_i{1-tanh((r_i-r)/\xi)}$ (dark dashed line) with $\xi$ = 23 nm. This is an approximation widely used in literature, roughly confirmed by microscopic calculations which show that the intervortex density of states for single band superconductors in the dirty limit is indeed practically negligible for fields below about half H$_{c2}$ \cite{Golubov94,Koshelev03,Samuely13}. Our data show that the tunneling conductance in-between vortices is more affected by the magnetic field than expectations for a single gap superconductor. The magnetic field increase of the intervortex density of states is pronounced (open blue points in Fig.\ 5c and Fig.\ 6). It is smaller than the increase found in the superconductors 2H-NbSe$_2$, 2H-NbS$_2$ and MgB$_2$ but also above the increase expected for a single gap s-wave superconductor (dark dashed line in Fig.\ 5c). When we apply the magnetic field along the basal plane, we find a smaller sized intervortex density of states (Fig.\ 5c).

We have measured H$_{c2}(T)$ using resistivity and susceptibility (Fig.\ 7), and compared results with available data in a sample with a larger mean free path\cite{Imai12}. A smaller mean free path leads to a shorter coherence length and hence to an increased H$_{c2}$(T), also in multigap superconductors, see for instance MgB$_2$\cite{Gurevich07,Gurevich04,Budko02,Budko05,Putti04}. Here, however, we do not observe such an increase in H$_{c2}$ and instead find the same result as in previous measurements on samples with a larger mean free path.

\begin{figure}
\includegraphics[width=0.45\textwidth]{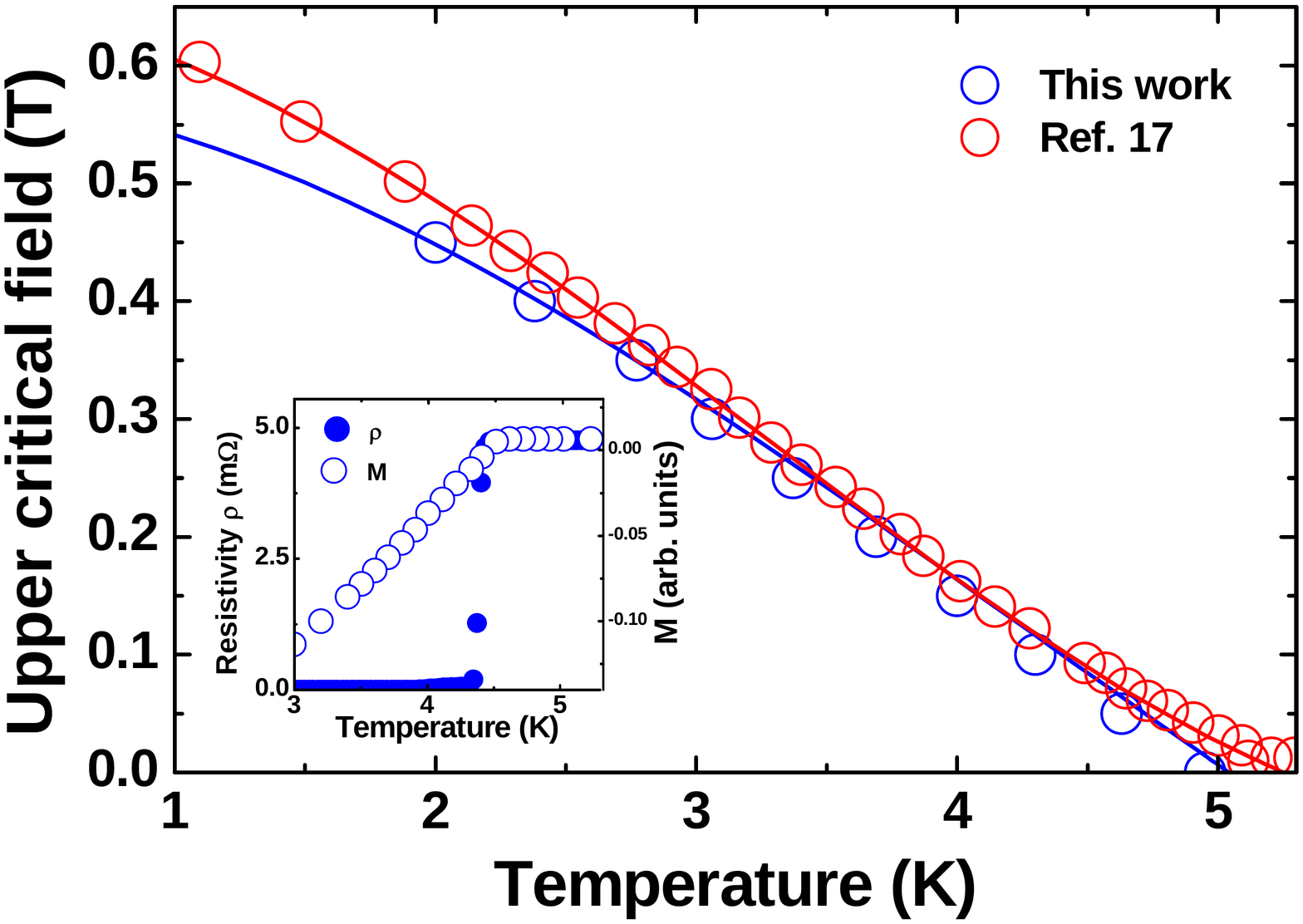}
\caption{a) H$_{c2}(T)$ along the c-axis in our sample of $\beta$-Bi$_2$Pd with a residual resistivity of 18 $\mu\Omega$cm (open blue points) and H$_{c2}(T)$ in sample of Ref.\protect\cite{Imai12} with a residual resistivity of 5 $\mu\Omega$cm (open red points). Lines are fits to each H$_{c2}(T)$ using the parameters explained in the text. Note that the upper critical field remains nearly the same, although the mean free path varies by a factor of four. Magnetic field is applied along the c-axis. In the inset we show the transition in resistivity and susceptibility in our sample at a magnetic field of 0.1 T. The transitions remain sharp and well defined throughout the phase diagram.}
\end{figure}

To explain this result, we have calculated $H_{c2}(T)$ using a multiband approach and the method described in Refs.\cite{Werthamer66,Shulga98,Suderow04b,Suderow05d,Tissen13} (see also Annex for details). We simplify the Fermi surface by using a two band description, with differing Fermi velocities $v_{F,i}$, electron-phonon coupling $\lambda_{ij}$ and electron scattering parameters $\tau_{i,j}$ (with subindices $i,j$ for each band). We assume that the intraband and interband electron-phonon coupling constants $\lambda_{ij}$ are connected together as $\lambda_{11}=\lambda_{21}$ and $\lambda_{22}=\lambda_{12}$. The superconducting gap has then the same value in both bands. This reproduces our zero field tunneling spectroscopy result (Fig.\ 2). Specific heat measurements made very recently in our samples are also compatible with a single superconducting gap\cite{SamuelyToBePublished}. For the electronic specific heat coefficient we use $\gamma$=13 $mJ/K^2mol$ and derive the partial densities of states for each band from the anisotropy in the Fermi velocity. We also correlate the values of the $\tau_{i,j}$ with the experimental values of the resistivities. To this end, we assume that the conductivities of each band are equal $\rho_1=\rho_2$ and we deduce the relaxation rate for each part of the Fermi surface using $\tau_{i}= \frac{\rho_i \gamma_i v_{Fi}^2}{V_{mol}}$. Where $V_{mol}$ is the molar volume, $\gamma$ the Sommerfeld coefficient of the specific heat and $\rho$ the resistivity.  Moreover, $\lambda_{12}/\lambda_{21}$ is given by the ratio of the density of states in both bands. Thus, taking $\gamma = 13\,mJ/K^2 mol$, we deduce $\gamma_i$ from the choice of the $\lambda_{i,j}$.  We also assumed that the relaxation rate is isotropic and that each $\tau_i$ is split into intraband and interband scattering, parametrized by an additional coefficient, $\alpha$, such that $\tau_{11}=\alpha\tau_{1}$, $\tau_{12}=(1-\alpha)\tau_{1}$. The introduction of interband scattering $\tau_{12}\ne 0$ is needed to decrease the sensitivity of H$_{c2}(T)$ to the mean free path $\ell$ and obtain similar values of $H_{c2}(T)$ when $\ell$ is decreased. In table I, we give the list of parameters values used to fit the two sets of data.

\begin{table*}[htbp]
\centering
		\begin{tabular}{ccccccccc}
			%\hline \hline
		 \multicolumn{8}{c}{Parameters for the calculation of H$_{c2}$(T)}\\
		\hline \hline
		Sample & ${\it v}_{F1}$ & ${\it v}_{F2}$ & $\gamma$ ($mJ/K^2mol)$  & $\rho$ ($\mu \Omega$ cm) & $T_c$ (K) & $\alpha$ & $\beta$  \\
			\hline \hline
			This work &  0.09 & 0.4 & 13 & 18 & 5.07 & 0.2 & 0.8 \\	
			Ref. Imai \protect\cite{Imai12} &  0.09 & 0.4 & 13 & 5 & 5.35 & 1 & 1 \\	
						\hline
			\end{tabular} 
		\caption{Parameters used to calculate the temperature dependence of the upper critical field along the c-axis in samples used in this work and in samples of Ref.\protect\cite{Imai12} (Fermi velocities are given in units of 10$^6$ m/s).}
	\label{tab:ParametersUpperCriticalField}
\end{table*}

We neglect interband defect scattering for the data of Ref.\cite{Imai12}. This provides an excellent fit to the $H_{c2}(T)$ data of Ref.\cite{Imai12} (Fig.\ 7). If we simply decrease the mean free path to try to fit H$_{c2}(T)$ in our sample, we find an increase of the upper critical field. Only when allowing for interband scattering, with $\tau_{ij}\neq0$ for $i,j\neq 0$, we find that H$_{c2}(T)$ does not increase when the mean free path is decreased. An intraband decrease of the mean free path invariably leads to a decreased coherence length and an increased upper critical field, but interband mixing can lead to similar values of the upper critical field. Introducing interband scattering, we obtain an excellent fit of our data, with, in particular, an upper critical field that does not increase with decreased mean free path(Fig.\ 7).

\section{Discussion}

The influence on the vortex lattice of a square crystal symmetry has been discussed in detail before. At low magnetic fields, neutron scattering studies of the vortex lattice in tetragonal superconductors such as TmNi$_2$B$_2$C and CeCoIn$_5$ show two vortex lattice domains\cite{Eskildsen01b,Eskildsen01,Eskildsen03,DeBeer07}. In other tetragonal nickel borocarbides, and in V$_3$Si, the square symmetry produces a transition between hexagonal and square vortex lattices when increasing the magnetic field above fields of the order of a Tesla. The current distribution around vortices is sensitive to non-local effects, which introduce a radial dependence in the vortex-vortex repulsion due to the shape of the Fermi surface\cite{Kogan97,Eskildsen01b,Eskildsen01,Sosolik03,Guillamon10}. In $\beta-$Bi$_2$Pd, the square electronic symmetry does not give transitions in the vortex lattice symmetry. But the vortex lattice is still locked to the crystal lattice. Thus, non-local effects remain, showing that the Fermi surface features are playing a prominent role in the orientation of the vortex lattice.

The lack of variation of H$_{c2}$ with mean free path points towards the influence of multiband Fermi surface on the mixed state properties.  In MgB$_2$, two-dimensional sheets with strong electron-phonon coupling are derived from the $\sigma$ electrons of B orbitals, whereas the three dimensional sheets are derived from $\pi$ band orbitals. Interband scattering is particularly small in MgB$_2$. Previous theoretical work remarked that a decrease in H$_{c2}$ might be observed by producing strong enough interband scattering\cite{Gurevich07,Gurevich04}. The experiments show, however, a strong decrease of T$_c$ with interband scattering, concomitant with a more isotropic superconducting gap and increased H$_{c2}$\cite{Budko02,Martinez03,Budko05,Putti04}.

$\beta-$Bi$_2$Pd has a Fermi surface with two nearly cylindrical sheets and two 3D sheets, derived from Pd 4d and Bi 6p states\cite{Shein12,Sakano15}. States from both Pd and Bi contribute to the density of states at the Fermi level, and in particular anisotropic $4d_{xy+yz}$ orbitals. The strong anisotropy of these orbitals favors interband scattering. With a single superconducting gap, the zero field T$_c$ is not strongly affected by interband scattering, yet H$_{c2}$ shows the peculiar behavior discussed here.

\begin{table}
	\centering
		\begin{tabular}{c|cc|cc}
			%\hline \hline
		& \multicolumn{2}{c}{FS parameters from $H_{c2}(T)$} & \multicolumn{2}{c}{$\Delta$} (meV)\\
		\hline \hline
			 &  ${\it v}_{F1}$ (10$^6$ m/s) & ${\it v}_{F2}$ (10$^6$ m/s) & $\Delta_>$ & $\Delta_<$ \\
			\hline
			MgB$_2$          & 0.29 & 0.9 & 7.1 & 2.2 \\	
			2H-NbS$_2$      & 0.155 & 3.1 & 0.97 & 0.53 \\
			2H-NbSe$_2$      & 0.055 & 1 & 1.2 & 0.75 \\
			$\beta$-Bi$_2$Pd & 0.09 & 0.4 & 0.75 & 0.75 \\
		\end{tabular} 
		\caption{Parameters used to account for $H_{c2}(T)$ of MgB$_2$, 2H-NbS$_2$, 2H-NbSe$_2$ and $\beta-$Bi$_2$Pd. The gap values $\Delta_>$ and $\Delta_<$ are obtained from \protect\cite{Eskildsen02,Eskildsen03,Guillamon08PRB,Guillamon08c,Suderow04b,Suderow05d,Tissen13,Martinez03}.}
	\end{table}

We now compare our results results on the deviation in the magnetic field dependence of intervortex density of states with respect to expectation for single gap superconductors in $\beta$-Bi$_2$Pd with known features of MgB$_2$, 2H-NbSe$_2$ and 2H-NbS$_2$ (table II) \cite{Eskildsen02,Eskildsen03,Guillamon08PRB,Guillamon08c,Suderow04b,Suderow05d,Tissen13,Martinez03}. MgB$_2$ is the compound where the two gaps are more separated in energy, being the larger gap a factor of three higher than the smaller gap. Moreover, the gap distribution is narrow around these two values providing two neat features in the energy dependence of the superconducting density of states\cite{Martinez03}. The latter is also found in 2H-NbS$_2$ although the ratio between the two gaps is somewhat smaller, around two\cite{Guillamon08c}. The same gap ratio is found in 2H-NbSe$_2$, although in this material there is a sizeable in-plane gap anisotropy with a wide distribution of gap values\cite{Guillamon08PRB,Johannes06,Rahn12,Rodrigo04PhysC}. This makes the vortex core overlap in 2H-NbSe$_2$ (Fig.\ 5) smaller than in 2H-NbS$_2$. On the other hand, the anisotropy in the Fermi velocity is stronger in 2H-NbS$_2$ and 2H-NbSe$_2$ than in MgB$_2$\cite{Eskildsen02,Eskildsen03,Guillamon08PRB,Guillamon08c,Suderow04b,Suderow05d,Tissen13}. Yet, the vortex core overlap is the highest in the latter. Thus, the strongest increase in vortex core overlap is produced by multigap superconductivity.

In conclusion, we have obtained atomically flat Bi surfaces in $\beta-$Bi$_2$Pd where we observe an isotropic superconducting gap and a hexagonal vortex lattice. By discussing tunneling spectroscopy, vortex lattice and H$_{c2}(T)$, we have shown that $\beta-$Bi$_2$Pd is a multiband superconductor with a single superconducting gap. Interband scattering precludes the usual increase of H$_{c2}$ with a decreased mean free path. The hexagonal vortex lattice orientation locks to the crystalline lattice. We conclude that the mixed phase of superconductors is strongly modified in multiband Fermi surface materials, even when the zero field superconducting density of states is not.

\begin{acknowledgments}
We wish to acknowledge the support of COLCIENCIAS Programa Doctorados en el Exterior Convocatoria 568-2012. This work was supported by the Spanish MINECO (FIS2011-23488 and MAT2011-27470-C02-02), by the Comunidad de Madrid through program NANOFRONTMAG-CM (S2013/MIT-2850) and by Axa Research Fund. We also acknowledge SEGAINVEX workshop of UAM, Banco Santander, Graphene Flagship (EU Grant Agreement No. 604391), COST MP1201 action, and in depth discussions with P. Samuely. We particularly acknowledge P.C. Canfield for setting up together with us our growth lab and teaching us details about crystal growth.
\end{acknowledgments}

\section{Annex I. Calculation of the upper critical field.}

Several papers provide methods to calculate the upper critical field $H_{c2}(T)$ in multiband superconductors\cite{Werthamer66,Shulga98,Suderow04b,Suderow05d,Tissen13,Gurevich07,Gurevich04,Putti04}. Here we follow Ref.\cite{Shulga98}, a microscopic calculation of $H_{c2}(T)$ taking into account defect scattering. The upper critical field $H_{c2}(T)$ is found by calculating the set of $\beta_i$ with largest values that solve the following equations:

\begin{eqnarray}
\varpi_i(n)=\omega_n+\pi T\sum_{j,m}(\lambda_{i,j}(m-n)+\nonumber \\
\delta_{mn}(\tau_{i,j}/2\pi T))\text{sgn}(w_{m})
\end{eqnarray}

\begin{eqnarray}
\Delta_i(n)=\pi T \sum_{j,m}[\lambda_{i,j}(m-n)-
\mu^*\delta_{ij}\theta(\omega_c-|{\omega_m}|)+\nonumber \\
\delta_{mn}(\tau_{i,j}/2\pi T)]\mathcal{X}_j(m)\Delta_j(m)
\end{eqnarray}

\begin{eqnarray}
\mathcal{X}_{i}(n)=(2/\sqrt{\beta_i})\int^\infty_0 d q e^{-q^2} tan^{-1}(q\sqrt{\beta_i}/(|\varpi_i(n)|+\nonumber \\
\frac{ig}{2}\mu_B H_{c2}\text{sgn}(\omega_n))
\end{eqnarray}

\begin{equation}
\beta_i=\frac{e}{2}H_{c2}v_{Fi}^2
\end{equation}

\begin{equation}
\lambda_{i,j}(n)=\int_0^\infty d\omega \omega \alpha^2_{i,j}\frac{F(\omega)}{\omega^2+\omega_n^2}
\end{equation}

$\omega_n$ are the Matsubara frequencies, $\omega \alpha^2_{i,j}F(\omega)$ is the electron-phonon coupling and $\Delta_i$ the Cooper pair wavefunction, $\tau_{i,j}$ the relaxation rate, $\lambda_{i,j}$ the electron phonon coupling constant and $v_{F,i}$ the Fermi velocity at a plane perpendicular to the magnetic field in each band $i$. The Fermi velocities discussed here are unrenormalized, i.e. as obtained without the pairing interactions. Their values eventually found in experiments (as quantum oscillation or photoemission) need to be renormalized by the electron-phonon interaction $\lambda_{i,j}$.

We see that the relevant parameters to describe mixture between different bands are the off-diagonal terms of matrices $\lambda_{i,j}$ and $\tau_{i,j}$. The superconducting order parameter is found by an equation of the same form of the BCS self-consistency gap equation (2). The first term of the equation between square brackets $[$ $]$ accounts for intraband and interband scattering due to electron-phonon interaction and defect scattering.

The bare Fermi velocity $v_{F,i}$ enters into the equation through the term $\mathcal{X}_j$, which also depends on the electron-phonon coupling and the interband and intraband scattering parameters $\lambda_{i,j}$ and $\tau_{i,j}$ through equation (1).

%\bibliography{LastBib_noTitle}

\end{document}